\documentclass[a4paper,12pt]{article}
\usepackage{amssymb}

\begin{document}

\title{Physical Significance of the Difference between \\
the Brans-Dicke Theory and General Relativity}
\author{A. Miyazaki \thanks{
Email: miyazaki@loyno.edu, miyazaki@nagasakipu.ac.jp} \vspace{3mm} \\
\textit{Department of Physics, Loyola University, New Orleans, LA 70118} \\
and \\
\textit{Faculty of Economics, Nagasaki Prefectural University} \\
\textit{Sasebo, Nagasaki 858-8580, Japan}}
\date{\vfill}
\maketitle

\begin{abstract}
The asymptotic behavior of the scalar field and its physical meaning are
discussed for $T=0$ and $T\neq 0$ for the large enough coupling parameter $%
\omega $. The special character of the Brans-Dicke theory is also discussed
for local and cosmological problems in comparison with general relativity
and the selection rules are introduced respectivley. The scalar field by
locally-distributed matter should exhibit the asymptotic behavior $\phi
=\left\langle \phi \right\rangle +O(1/\omega )$ because of the presence of
cosmological matter. The scalar field of a proper cosmological solution
should have the asymptotic form $\phi =O(\rho /\omega )$ and should converge
to zero in the continuous limit $\rho /\omega \rightarrow 0$.\newline
\ \newline
\textbf{PACS number(s): 04.50.+h, 98.80.-k }
\end{abstract}

\newpage

\section{Introduction}

It seems that Einstein's general relativity has increasingly obtained its
numerical validity by many experimental and observational tests.
Nevertheless, on the other hand, much efforts for scalar-tensor theories of
gravitation also continue for a long time. We have some historical or
fundamental bases on which we believe that there should exist some kinds of
scalar fields as the gravitational field.

The Brans-Dicke theory \cite{1)} is the prototype of such scalar-tensor
theories of gravitation, and the gravitational field is described by the
metric tensor $g_{\mu \nu }$ of the Riemannian manifold and the
non-minimally coupled scalar field $\phi $ on that manifold, which
represents the spacetime-varying gravitational ''constant''. The field
equations of the Brans-Dicke theory are obtained by the similar variational
method as the Einstein theory, and are given as following in our sign
conventions: 
\begin{eqnarray}
R_{\mu \nu }-\frac{1}{2}Rg_{\mu \nu } &=&\frac{8\pi }{c^{4}\phi }T_{\mu \nu
}-\frac{\omega }{\phi ^{2}}\left( \phi _{,\,\mu }\phi _{,\,\nu }-\frac{1}{2}%
g_{\mu \nu }\phi _{,\,\lambda }\phi ^{,\,\lambda }\right)  \nonumber \\
&&-\frac{1}{\phi }(\phi _{,\,\mu ;\,\nu }-g_{\mu \nu }\square \phi )\,,
\label{e1}
\end{eqnarray}
\begin{equation}
\square \phi =-\frac{8\pi }{(3+2\omega )c^{4}}T\,,  \label{e2}
\end{equation}
where $T_{\mu \nu }$ is the energy-momentum tensor of matter and $\omega $
is the coupling parameter of the scalar field.

As mentioned in many literatures (see, for example, \cite{2)}), when the
coupling parameter $\omega $ is large enough, the scalar field and the field
equations of gravitation have the following approximate form: 
\begin{equation}
\phi =\left\langle \phi \right\rangle +O(1/\omega )\,,  \label{e3}
\end{equation}
\begin{equation}
R_{\mu \nu }-\frac{1}{2}Rg_{\mu \nu }=\frac{8\pi }{c^{4}\phi }T_{\mu \nu
}+O(1/\omega )\,,  \label{e4}
\end{equation}
and in the limit of infinity ($\omega \rightarrow \infty $), the scalar
field $\phi $ converges to constant $\left\langle \phi \right\rangle $, thus
the field equations of gravitation coincide completely with those of general
relativity by replacing $\phi $ with Newton's gravitational constant $%
G\equiv \left\langle \phi \right\rangle ^{-1}$.

Recently, however, some authors \cite{3)}, \cite{4)} reported that these
discussions are generally not right when the contracted energy-momentum
tensor $T=T_{\mu }^{\;\mu }$ vanishes. According to Banerjee and Sen \cite
{3)}, in this situation $T=0$, asymptotic behavior of the scalar field
becomes 
\begin{equation}
\phi =\left\langle \phi \right\rangle +O(1/\sqrt{\omega })  \label{e5}
\end{equation}
when the coupling parameter is large enough. In the limit of infinity,
though the scalar field definitely converges to constant, the second term of
the right-hand side of Eq.(\ref{e1}) remains nonvanishing and the field
equations of the Brans-Dicke theory do not coincide with those of the
Einstein theory with the same energy-momentum tensor $T_{\mu \nu }$. As for
such examples of exact solutions, see Refs of \cite{3)}, \cite{4)}. They say
that \emph{the condition} $T\neq 0$ \emph{is both necessary and sufficient
for the Brans-Dicke solutions to yield the corresponding solutions of
general relativity with the same} $T_{\mu \nu }$ \emph{in the infinite }$%
\omega $ \emph{limit}.

However, this theorem is not true as indicated by Faraoni \cite{4)} with a
counterexample \cite{5)}. Faraoni gave a rigorous mathematical proof to the
asymptotic behavior Eq.(\ref{e5}) by discussing the conformal invariance of
the Brans-Dicke theory when $T=0$. He insists only that \emph{the
Brans-Dicke solutions with} $T=0$ \emph{generically fail to reduce to the
corresponding solutions of general relativity when} $\omega \rightarrow
\infty $.

In this article we will add another counterexample to the above theorem \cite
{3)}, and will discuss generally the physical meaning of the relationship
between the Brans-Dicke theory and the Einstein theory in the cases $T=0$
and $T\neq 0$ in contexts of local or cosmological problems. Rather, going
back to the original motivation of the Brans-Dicke theory, the Machian point
of view, we realize that both are different theories of gravitation, and
that the Brans-Dicke theory need hardly reduce to the Einstein theory when $%
\omega \rightarrow \infty $ except local problems.

\section{Physical Meaning of the Asymptotic Behavior}

Let us discuss the asymptotic behavior of the scalar field when $T=0$. An
order of magnitude estimate by Banerjee and Sen \cite{3)} is more
appropriate to understand its physical meaning. When $T=0$, we obtain from
Eq.(\ref{e2}) 
\begin{equation}
\square \phi =0\,,  \label{e6}
\end{equation}
and get from the trace of Eq.(\ref{e1}) 
\begin{equation}
R=-\frac{8\pi }{c^{4}\phi }T-\frac{\omega }{\phi ^{2}}\phi _{,\,\lambda
}\phi ^{,\,\lambda }-\frac{3}{\phi }\square \phi \,.  \label{e7}
\end{equation}
It is easy to see the asymptotic behavior Eq.(\ref{e5}) of the scalar field
from this equation when $T=0$. However, remember we assume tacitly that the
scalar field $\phi $ converges to constant in the infinite $\omega $ limit
and the scalar curvature $R$ does not depend on $\omega $, both of which do
not seem to be obvious. Moreover, the Minkowski space with $T=0$ and $R=0$
has only the constant scalar $\left\langle \phi \right\rangle $, which is
independent of the coupling parameter $\omega $. It is to be remarked that a
solution satisfying 
\begin{equation}
\frac{\omega }{\phi ^{2}}\left( \phi _{,\,\mu }\phi _{,\,\nu }-\frac{1}{2}%
g_{\mu \nu }\phi _{,\,\lambda }\phi ^{,\,\lambda }\right) =0
\end{equation}
is only $\phi =const$ \cite{4)}. Therefore, all Einstein spaces ($R_{\mu \nu
}=0$) with constant $\phi $ are exceptions for the statement by Banerjee and
Sen, or by Faraoni.

When $T\neq 0$ the asymptotic form of the field equation (\ref{e2}) becomes 
\begin{equation}
\square \phi =O(1/\omega )\,,  \label{e8}
\end{equation}
and we observe the well-known asymptotic behavior Eq.(\ref{e3}) of the
scalar field. We, however, should strictly read Eq.(\ref{e8}) as $\square
\phi =O(T/\omega )$, or for simplicity 
\begin{equation}
\square \phi =O(\rho /\omega )  \label{e9}
\end{equation}
for dust matter.

Now we can understand the physical meaning of the difference of the
asymptotic behavior of the scalar field. The Brans-Dicke theory includes
originally the coupling parameter $\omega $\ in the right-hand side of Eq.(%
\ref{e2}) and in the second term of the right-hand side of Eq.(\ref{e1}).
When $T\neq 0$, the dependence of $\omega $\ in the scalar field comes
fundamentally from the coupling parameter $\omega $\ in the right-hand side
of Eq.(\ref{e2}), and the second term of the right-hand side of Eq.(\ref{e1}%
) vanishes in the infinite $\omega $\ limit. When $T=0$, the right-hand side
of Eq.(\ref{e2}) vanishes and the dependence of $\omega $\ comes
fundamentally from $\omega $\ in the second term of the right-hand side of
Eq.(\ref{e1}).\ The Brans-Dicke scalar field has finite indefiniteness $\phi
_{V}(x^{\mu })$ which satisfies the d'Alembertian equation (\ref{e6}) even
when matter does not exist ($T=0$). This scalar field $\phi _{V}(x^{\mu })$
is constrained by another field equation (\ref{e1}), and thus it has the
dependence of $\omega $ like Eq.(\ref{e5}). The scalar field $\phi
_{V}(x^{\mu })$, which behaves like a source of the gravitational field $%
g_{\mu \nu }$ in Eq.(\ref{e1}), has no material origin. The constant part $%
\left\langle \phi \right\rangle $ itself is a special case of this scalar
field without material origin. To the contrary, the asymptotic behavior of
the scalar field with material origin is determined by its field equation (%
\ref{e2}) with the source term.

In general, when matter exists ($T\neq 0$), the scalar field includes a part
given by matter and indefiniteness $\phi _{V}(x^{\mu })$ for $T=0$, and its
asymptotic behavior for large $\omega $\ becomes 
\begin{equation}
\phi =\left\langle \phi \right\rangle +O(1/\sqrt{\omega })+O(1/\omega )\,.
\label{e10}
\end{equation}
It is clear that the term of $O(1/\sqrt{\omega })$ is dominant when the
coupling parameter $\omega $ is large enough. Therefore the Brans-Dicke
theory fails to yield general relativity in the infinite $\omega $\ limit.
This situation produces possible counterexamples to the theorem proposed by
Banerjee and Sen. An example cited by them to reinforce the theorem, a
closed vacuum ($T=0$) Friedmann-Robertson-Walker solution with cosmological
constant $\Lambda $ \cite{6)}, \cite{7)}, should rather be included here
because of $\square \phi =2\Lambda \phi /(2\omega +3)\neq 0$ though $T=0$.

\section{Local Problems}

When we consider the difference between the Brans-Dicke theory and the
Einstein theory, we should realize the difference between local problems
(with locally-distributed matter) and cosmological (or global) problems.
Brans and Dicke \cite{1)} also comment, in discussing a Schwarzschild
solution in their theory, that we premise the existence of distant matter in
the universe. Our universe always exists, and in the Brans-Dicke theory we
need discuss local problems in the presence of cosmological matter, which
supports the gravitational ''constant''. In the framework of general
relativity, which has \emph{a priori} gravitational constant, we do not
consider our environment of the universe and discuss purely the local
gravitational field with an asymptotically-flat boundary condition.

It might be very difficult to solve globally all configurations of matter in
the universe in the Brans-Dicke theory. However, it seems to be a good
enough approximation to divide the two side, local and cosmological
problems, because our universe is huge enough. We can discuss individual
problems of locally-distributed matter with an asymptotically-flat boundary
condition. To do so, we need accept two postulates; We use an experimental
value of gravitational constant for the constant scalar field $\left\langle
\phi \right\rangle =1/G$, and request that the local scalar field $\phi $
also converges to $\left\langle \phi \right\rangle $ at the distant enough
region ($r\rightarrow \infty $). Moreover, we adopt a selection rule, which
is derived by the presence of cosmological matter in the universe. Let us
consider the static spherically symmetric vacuum solution \cite{1)} in the
Brans-Dicke theory (only scalar part): 
\begin{equation}
\phi =\phi _{0}\left( \frac{1-B/r}{1+B/r}\right) ^{-C/\sigma }  \label{e11}
\end{equation}
where $B=(M/2C^{2}\phi _{0})[(2\omega +4)/(2\omega +3)]^{1/2}$, $\sigma
=[(C+1)^{2}+C(1-\frac{1}{2}\omega C)]^{1/2}$, and $C$ is arbitrary constant.
It is obvious that this solution converges to the constant $\phi _{0}$ in
the infinite $\omega $\ limit. In a case of arbitrary constant $C$
(independent of $\omega $), the asymptotic form of this solution becomes $%
\phi =\left\langle \phi \right\rangle +O(1/\sqrt{\omega })$ \cite{3)}, which
means that Eq.(\ref{e11}) does not produce the corresponding solution of
general relativity, the Schwarzschild solution. We need suppress
indefiniteness of solutions with no material origin. We can not accomplish
enough this work by boundary conditions. Our selection rule act to choose a
proper solution (or solutions), of which asymptotic behavior is Eq.(\ref{e3}%
) with material origin, and which yields a corresponding solution of general
relativity. Because the corresponding exact and global Brans-Dicke solution
is originally generated by the nonvanishing energy-momentum tensor ($T_{\mu
\nu }\neq 0$) by locally-distributed and cosmological matter in the
universe. We can forget the effect of the presence of cosmological matter if
we set the two postulates for local problems in the Brans-Dicke theory. In
this standpoint, general relativity is the self-complete approximate-theory
of gravitation as it needs no additional postulates.

If we take a choice $C=-1/(2+\omega )$, the equation (\ref{e11}) behaves
asymptotically as Eq.(\ref{e3}) for the large enough $\omega $\ and the
whole solution becomes identical with the Schwarzschild solution of the
Einstein theory for $\omega \rightarrow \infty $ \cite{1)}, \cite{6)}.
Another choice $C=-1/2\omega $ is also available \cite{4)}. However, it is
to be remarked that this $\omega $\ has no meanings though the same letter $%
\omega $\ is used. This is not the coupling parameter $\omega $\ derived
from the right-hand side of Eq.(\ref{e2}) or the second term of the
right-hand side of Eq.(\ref{e1}). If we regard Eq.(\ref{e11}) as the
approximate and local solution\ for a point-mass $M$\ at the origin and
cosmological matter in the universe, we may be able to interpret this $%
\omega $\ as the real coupling parameter derived from the right-hand side of
Eq.(\ref{e2}) with source matter. Though this is a conjecture, let us adopt
as a postulate. For local problems we should forgive only solutions of which
the asymptotic behavior is $\phi =\left\langle \phi \right\rangle
+O(1/\omega )$ even when $T=0$. Thus we can restrict indefiniteness and
select proper solutions with material origin. For local problems they are
equivalent to each other to have material origin, to behave asymptotically
as Eq.(\ref{e3}) for the large enough $\omega $, and to converge to the
corresponding solution of general relativity in the infinite $\omega $
limit. There exists arbitrariness of forgiven solutions, for example a
choice of $C$, and remains controversial. They give different solutions for
finite values of $\omega $.

The contracted energy-momentum tensor $T$ is zero for these solutions.
Nevertheless the corresponding solution of general relativity is produced in
the infinite $\omega $\ limit. However, this should not be regarded as a
counterexample to the proposition by Faraoni \cite{4)}. This proposition
should be understood to stand for exact solutions with exact $T=0$. In his
standpoint, the statement $T=0$ means that there exist completely no other
matter in the universe. He treats local problems for locally-distributed
matter as whole cosmological problems. To the contrary, though we use the
statement $T=0$ in these solutions, this is a local approximation for local
problems and actually the complete global contracted energy-momentum tensor
does not vanish ($T\neq 0$) because of the presence of other cosmological
matter in the universe. If we discuss locally-distributed matter in
otherwise empty space, the scalar field $\phi $\ should not have the
constant scalar field $\left\langle \phi \right\rangle $. This is the
important keynote to understand the Brans-Dicke theory true. It is
meaningless that we consider strictly the situation in which matter does not
exist, or vacuum space in the Brans-Dicke theory.

\section{Cosmological Problems}

Next we consider cosmological problems to make clear further the essence of
the Brans-Dicke theory. Let us discuss first the Brans-Dicke flat solution 
\cite{1)} for the homogeneous and isotropic universe. Assuming the initial
conditions 
\begin{equation}
\phi =a=0\,;\;\;t=0\,,  \label{e12}
\end{equation}
it is given as 
\begin{equation}
ds^{2}=-dt^{2}+a^{2}(t)[d\chi ^{2}+\chi ^{2}(d\theta ^{2}+\sin ^{2}\theta
d\varphi ^{2})]\,,  \label{e13}
\end{equation}
\begin{equation}
\phi =\phi _{0}(t/t_{0})^{r}\,,\;\;a=a_{0}(t/t_{0})^{q}\,,\;\;\rho
a^{3}=\rho _{0}a_{0}^{3}\,,  \label{e14}
\end{equation}
with 
\begin{equation}
r=2/(4+3\omega )\,,\;\;q=(2+2\omega )/(4+3\omega )\,,  \label{e15}
\end{equation}
and 
\begin{equation}
\phi _{0}=4\pi \lbrack (4+3\omega )/(3+2\omega )c^{2}]\rho _{0}t_{0}^{2}\,,
\label{e16}
\end{equation}
where $\rho _{0}$ is the present mass density. For the large coupling
parameter $\omega $, it is easy to observe that the scalar field $\phi $\ of
this solution behaves like Eq.(\ref{e3}) \cite{6)}. If the mass density $%
\rho _{0}$ decreases to zero, the scalar field $\phi $\ also converges to
zero, and this situation is suitable for the material origin of the scalar
field. It is well-known that this solution reduces to the Einstein-de Sitter
universe of general relativity in the infinite $\omega $\ limit.\ However,
what does the constant value $\left\langle \phi \right\rangle =6\pi \rho
_{0}t_{0}^{2}/c^{2}$ mean? In the infinite $\omega $\ limit, the coupling
between the scalar field and matter vanishes. Why does the mass density $%
\rho _{0}$ appear in the constant scalar field $\left\langle \phi
\right\rangle \,$? Which value of the density should we take? This situation
is rather strange as to the material origin. After all, this constant scalar
field $\left\langle \phi \right\rangle $ seems to be merely constant which
has no material origin.

O'Hanlon and Tupper \cite{8)} solution for a vacuum, spatially flat
Friedmann-Robertson-Walker spacetime has the asymptotic behavior $\phi
=\left\langle \phi \right\rangle +O(1/\sqrt{\omega })$ \cite{6)}, which
means that this solution has no material origin. This solution also has the
constant scalar field for $\omega \rightarrow \infty $. Nariai \cite{9)}
flat solution with a perfect fluid has the asymptotic behavior Eq.(\ref{e5})
for $T=0$ (radiation), and Eq.(\ref{e3}) for $T\neq 0$ (matter) \cite{3)}.
In both cases the scalar field converges to the constant $\phi _{0}$ in the
infinite $\omega $\ limit.

The following cosmological solution \cite{10)}, \cite{11)} gives both a
counterexample to the theorem by Banerjee and Sen \cite{3)} and an
interesting example as to the material origin of cosmological solutions.
Dehnen and Obreg\'{o}n says that this model has no analogy in general
relativity \cite{10)}, but this is not adequate \cite{12)}. The Brans-Dicke
theory has a particular closed solution for the homogeneous and isotropic
universe with dust ($T=\rho c^{2}$), satisfying $a(t)\phi (t)=const$, 
\begin{equation}
ds^{2}=-dt^{2}+a^{2}(t)[d\chi ^{2}+\sin ^{2}\chi (d\theta ^{2}+\sin
^{2}\theta d\varphi ^{2})]\,,  \label{e17}
\end{equation}
\begin{equation}
\phi (t)=-[8\pi /(3+2\omega )c^{2}]\rho (t)t^{2}\,,  \label{e18}
\end{equation}
\begin{equation}
a(t)=\left[ -2/(2+\omega )\right] ^{1/2}\,t\,,  \label{e19}
\end{equation}
\begin{equation}
2\pi ^{2}a^{3}(t)\rho (t)=M\,,  \label{e20}
\end{equation}
with 
\begin{equation}
\omega <-2\,,\;\;\;G(t)M/c^{2}a(t)=\pi \,,  \label{e21}
\end{equation}
where the gravitational ''constant'' $G=(4+2\omega )/(3+2\omega )\phi $ and $%
M$ is the total mass of the universe. The scalar field has obviously the
asymptotic behavior $O(1/\omega )$, but does not have the constant value $%
\left\langle \phi \right\rangle $\ in the infinite $\omega $\ limit ($\omega
\rightarrow -\infty $).\ The expansion parameter $a$\ also has the $\omega $%
-dependence, which means the scalar curvature $R$ itself has the $\omega $%
-dependence.

Let us write down the nonvanishing components of the field equations (\ref
{e1}) and (\ref{e2}) for the metric Eq.(\ref{e17}) to discuss the details of
the asymptotic behavior: 
\begin{equation}
2\dot{a}\ddot{a}+\dot{a}^{2}+1=-\frac{8\pi }{(3+2\omega )c^{2}}\frac{%
a^{2}\rho }{\phi }-\frac{1}{2}\omega a^{2}\left( \frac{\dot{\phi}}{\phi }%
\right) ^{2}+a\dot{a}\left( \frac{\dot{\phi}}{\phi }\right) \,,  \label{e22}
\end{equation}
\begin{equation}
\frac{3}{a^{2}}\left( \dot{a}^{2}+1\right) =\frac{16\pi (1+\omega )}{%
(3+2\omega )c^{2}}\frac{\rho }{\phi }+\frac{\omega }{2}\left( \frac{\dot{\phi%
}}{\phi }\right) ^{2}+\frac{\ddot{\phi}}{\phi }\,,  \label{e23}
\end{equation}
\begin{equation}
\ddot{\phi}+3\frac{\dot{a}}{a}\dot{\phi}=\frac{8\pi }{(3+2\omega )c^{2}}\rho
\,,  \label{e24}
\end{equation}
where a dot denotes the derivative with respect to $t$. For the small enough
coupling parameter ($\omega \ll -1$), we can estimate the order of each
terms by means of the solution, for example, 
\begin{equation}
\frac{8\pi }{(3+2\omega )c^{2}}\frac{a^{2}\rho }{\phi }\sim O(1/\omega )\,,
\label{e25}
\end{equation}
\begin{equation}
\frac{1}{2}\omega a^{2}\left( \frac{\dot{\phi}}{\phi }\right) ^{2}\sim
O(1)\,,  \label{e26}
\end{equation}
\begin{equation}
a\dot{a}\left( \frac{\dot{\phi}}{\phi }\right) \sim O(1/\omega )\,,
\label{e27}
\end{equation}
\begin{equation}
a^{2}\frac{\ddot{\phi}}{\phi }\sim O(1/\omega )\,.  \label{e28}
\end{equation}
Remark that the mass density $\rho $ is given and does not depend on $\omega 
$, and so $M$ has the $\omega $-dependence derived from $a$. The term Eq.(%
\ref{e26}), which is the contribution from the second term of the right-hand
side of Eq.(\ref{e1}), remains nonvanishing even though the scalar field has
the asymptotic behavior $O(1/\omega )$.

If we put $\lambda (t)\equiv -(\omega /2)(\dot{\phi}/\phi )^{2}$ and $\kappa
(t)\equiv 8\pi /c^{4}\phi $, we get from Eq.(\ref{e22}) and Eq.(\ref{e23}) 
\begin{equation}
2\dot{a}\ddot{a}+\dot{a}^{2}-\lambda a^{2}+1=O(1/\omega )\,,  \label{e29}
\end{equation}
\begin{equation}
\frac{3}{a^{2}}\left( \dot{a}^{2}+1\right) +\lambda =\kappa \rho
c^{2}+O(1)\,.  \label{e30}
\end{equation}
Thus we obtain in the infinite $\omega $\ limit ($\omega \rightarrow -\infty 
$) 
\begin{equation}
\lambda a^{2}=1\,,\;\;\;\kappa \rho c^{2}a^{2}=4\,.  \label{e31}
\end{equation}
\newline
If we regard $\lambda $ as the cosmological ''constant'', these relations
are similar to those of the static Einstein universe with negligible
pressure in general relativity except the difference of the radius of the
universe in $\sqrt{2}$ which is derived from the opposite sign of $\lambda $
in Eq.(\ref{e30}). In the infinite $\omega $\ limit, the expansion parameter 
$a$\ reduces to zero, but this is due to the initial condition $a=0$ at $t=0$
\cite{12)}. There exists a discrete and isolated limit at $\omega
\rightarrow -\infty $. For the finite $\omega $ ($\omega <-2$) we observe 
\begin{equation}
\lambda (t)a^{2}(t)=\omega /(2+\omega )\,,\;\;\;\kappa (t)\rho
(t)c^{2}a^{2}(t)=4\,,  \label{e32}
\end{equation}
\newline
and so the effective cosmological ''constant'' $\lambda (t)$ decreases
rapidly as the universe expands.

It is remarkable that the scalar field $\phi $ converges to zero for both
cases in which the mass density $\rho $\ goes to zero, and in which the
coupling parameter $\omega $ goes continuously to the infinity ($\omega
\rightarrow -\infty $). This situation is rather preferable for the material
origin of the scalar field. It means that the combination of $\rho /\omega $
plays an important role there.

Now we need consider the correspondence between the Brans-Dicke theory and
general relativity in combinations of $\left\langle \phi \right\rangle \neq
0 $ or $\left\langle \phi \right\rangle =0$, and $O(1/\omega )$ or\ $O(1/%
\sqrt{\omega })$. Let us put for abbreviation 
\begin{equation}
A_{\mu \nu }\equiv \frac{\omega }{\phi ^{2}}\left( \phi _{,\,\mu }\phi
_{,\,\nu }-\frac{1}{2}g_{\mu \nu }\phi _{,\,\lambda }\phi ^{,\,\lambda
}\right) \,,  \label{e33}
\end{equation}
\begin{equation}
B_{\mu \nu }\equiv \frac{1}{\phi }(\phi _{,\,\mu ;\,\nu }-g_{\mu \nu
}\square \phi )\,,  \label{e34}
\end{equation}
\begin{equation}
C\equiv \frac{8\pi }{c^{4}\phi }\,.  \label{e35}
\end{equation}
\newline
We can summarize orders of magnitude of each terms in $\omega $ or $\sqrt{%
\left| \omega \right| }$ as following: \newline
case (i) $\phi =\left\langle \phi \right\rangle +O(1/\omega )$, $%
\left\langle \phi \right\rangle \neq 0\,$, 
\begin{equation}
A_{\mu \nu }\sim O(1/\omega )\,,\;B_{\mu \nu }\sim O(1/\omega )\,,\;C\sim
O(1)\,,
\end{equation}
\newline
case (ii) $\phi =\left\langle \phi \right\rangle +O(1/\sqrt{\left| \omega
\right| })$, $\left\langle \phi \right\rangle \neq 0\,$, 
\begin{equation}
A_{\mu \nu }\sim O(1)\,,\;B_{\mu \nu }\sim O(1/\sqrt{\left| \omega \right| }%
)\,,\;C\sim O(1)\,,
\end{equation}
\newline
case (iii) $\phi =O(1/\omega )\,$, 
\begin{equation}
A_{\mu \nu }\sim O(\omega )\,,\;B_{\mu \nu }\sim O(1)\,,\;C\sim O(\omega )\,,
\end{equation}
\newline
case (iv) $\phi =O(1/\sqrt{\left| \omega \right| })\,$, 
\begin{equation}
A_{\mu \nu }\sim O(\omega )\,,\;B_{\mu \nu }\sim O(1)\,,\;C\sim O(\sqrt{%
\left| \omega \right| })\,.
\end{equation}
\newline
These results are derived on the assumption that the metric tensor $g_{\mu
\nu }$ converges to a nonvanishing function\ in the infinite $\omega $
limit. If $g_{\mu \nu }$ has other $\omega $-dependence which does not
satisfy this assumption, we need another individual analysis for the
specific solution, and the results seems to become different from the above.
Even for local problems we can not deny this possibility. However, it is
likely that $g_{\mu \nu }$ converges to a nonvanishing function\ in the
infinite $\omega $ limit if the scalar field $\phi $ converges to $%
\left\langle \phi \right\rangle $ $\neq 0$ for local problems. Anyhow, it is
common that the Brans-Dicke solutions fail to reduce to the corresponding
solutions of general relativity when $\left| \omega \right| \rightarrow
\infty $.

\section{Discussions}

Should the Brans-Dicke theory reduce to general relativity in the infinite $%
\omega $\ limit? No longer, its statement seems to be a preconception. It is
true that general relativity goes to the Newtonian theory of gravitation in
the weak field approximation ($GM/Rc^{2}\ll 1$). The fact that both general
relativity and the Newtonian theory have the common parameter, Newton's
gravitational constant $G$, makes it possible. However, the Brans-Dicke
theory and general relativity do not have a common parameter each other. The
infinite limit of the coupling parameter $\omega $ is ambiguous.\ After all,
it is natural to realize that the Brans-Dicke theory is a different theory
of gravitation from general relativity and need not reduce to it in the
infinite $\omega $\ limit. The differences of two theories are rather
essential in the physical meaning.

There exist at least three standpoints; First, general relativity is the
complete classical theory of gravitation. Second, the Brans-Dicke theory is
complete and is applicable to even purely local problems. Third, the
Brans-Dicke theory with some selection rules produces physically reasonable
solutions.

The first standpoint is the simplest and the most real, even though we can
not understand the origin of the gravitational constant $G$ and cannot help
accepting its value \emph{a priori}. The scalar field does not exist as the
gravitational field. We do not need the redundant Brans-Dicke theory.
General relativity is valid completely not only for local problems but also
for cosmological problems. The present experimental and observational tests
strongly support this standpoint.

In the second standpoint, the opposite extreme to the first, we can freely
apply the Brans-Dicke theory to all kinds of problems and formally obtain
their solutions if possible. We can discuss even a vacuum space. It may be
the Minkowski space. May a particle show the inertial property in this
space? We may set exactly asymptotically-flatness as a boundary condition
for locally-distributed matter. However, we encounter a serious difficulty
owing to ambiguity with $\square \phi =0$. This ambiguity is not avoidable
as long as we consider vacuum ($T=0$) even if we suppose specific boundary
conditions for the scalar field. We obtain different solutions with the
corresponding solutions of general relativity with the same energy-momentum
tensor in the infinite $\omega $\ limit for $T=0$

The third standpoint set the presence of matter forth as a premise; This is
the Machian point of view. We should clearly distinguish between local and
cosmological problems. Cosmology has a special situation in physics. We
always live in this universe and can not test alternatives. We cannot help
discussing every physical phenomena in this environment. We need adopt
selection rules for local and cosmological problems respectively. For local
problems, we restrict to solutions of which \emph{the scalar field has the
asymptotic behavior }$\phi =\left\langle \phi \right\rangle +O(1/\omega )$.
This selection rule is a reflection of the presence of cosmological matter
in the universe and suppresses ambiguity with $\square \phi =0$. Moreover, 
\emph{the cosmological background gives the finite reasonable scalar field }$%
\left\langle \phi \right\rangle $, that is, the gravitational ''constant'' $G
$ at the present time. Thus we can handle individual local problems with
asymptotically-flat boundary conditions for the metric tensor $g_{\mu \nu }$%
\ and the boundary condition for the scalar field $\phi $ ($\phi \rightarrow
\left\langle \phi \right\rangle $ as $r\rightarrow \infty $)\ in the
Brans-Dicke theory without considering our environment. This is an extremely
good approximation in this universe. We may discuss the space-varying $G$ by
locally-distributed matter in the ''empty'' space.

For cosmological problems, we adopt only a solution which behaves
asymptotically like $\phi =O(\rho /\omega )$. If the inertial properties are
determined completely by the presence of matter in the universe, the scalar
field should not include the constant $\left\langle \phi \right\rangle $
which has no material origin. The scalar field $\phi $\ should\ converge to
zero when the mass density $\rho $ decreases to zero in a proper
cosmological model. The scalar field should also converge to zero when the
coupling parameter $\left| \omega \right| $\ diverges to infinity and the
connection between the scalar field and matter vanishes. The mass density $%
\rho $ and the coupling parameter $\omega $\ are closely connected as the
source term of Eq.(\ref{e2}) and so we can combine two conditions. Thus 
\emph{the cosmological scalar field should converge to zero in the
continuous limit }$\rho /\omega \rightarrow 0$\emph{.} This situation is
crucial for considering the difference between the Brans-Dicke theory and
general relativity. The Minkowski space which has the constant scalar field
is excluded as a proper solution. A cosmological solution which fails to
reduce to the corresponding solution of general relativity is rather
physically reasonable. The constant scalar $\left\langle \phi \right\rangle $%
\ may be derived from the contribution of quantum corrections. However, this
contribution should be classically renormalized to the mass density because
the inertial-frame dragging is dominated completely by the distribution of
the mass density itself \cite{13)}.

All proper solutions for local problems in the third standpoint reduce to
the corresponding solutions of general relativity in the infinite $\omega $\
limit (, fixing cosmological part), so general relativity is the complete
approximate-theory of gravitation in this standpoint. Rather, it may be more
fundamental that we should adopt this fact as a postulate for local
problems. General relativity is effective enough for local problems and for
a small period for which the universe is quasi-static and the gravitational
''constant'' is constant enough.

It is a matter of course that experimental and observational tests should
finally determine which theory (, including other extended scalar-tensor
theories) is true and which standpoint is appropriate. After all, the
essential difference may appear only in cosmological problems owing to the
experimentally established large coupling parameter $\omega $. We have not
known an exact solution for our universe yet.\newline
\newline
\textbf{Acknowledgment}

The author is grateful to Professor Carl Brans for helpful discussions and
his hospitality at Loyola University (New Orleans) where this work was done.
He would also like to thank the Nagasaki Prefectural Government for
financial support.

\end{document}